\documentstyle[prl,aps,epsf,twocolumn]{revtex}
\draft
\begin{document}

\wideabs{
\title{A quasi-pure Bose-Einstein condensate immersed in a Fermi sea}
\author{F.\,Schreck, L.\,Khaykovich,  K.\, L.\,Corwin, G. Ferrari$^*$, T. Bourdel, J. Cubizolles, and C.\,Salomon}
\address{Laboratoire Kastler Brossel, Ecole Normale Sup\'erieure, 24 rue Lhomond, 75231 Paris CEDEX 05, France\\
$^*$ :LENS-INFM, Largo E. Fermi 2, Firenze 50125 Italy. }
\date{\today}
\maketitle

\begin{abstract}
We report the observation of co-existing Bose-Einstein condensate and Fermi gas in a
magnetic trap. With a very small fraction of thermal atoms, the $^{7}$Li condensate is
quasi-pure and in thermal contact with a $^{6}$Li Fermi gas. The lowest common temperature
is $0.28\,\mu$K $\simeq 0.2(1)\,T_{\rm C}= 0.2(1)\,T_{\rm F}$ where $T_{\rm C}$ is the BEC
critical temperature and $T_{\rm F}$ the Fermi temperature. Behaving as an
 ideal gas in the radial trap dimension, the condensate is one-dimensional.
\end{abstract}

\pacs{PACS numbers: 05.30.Fk, 05.30.Jp, 03.75.-b, 05.20.Dd, 32.80.Pj } }

 Bose-Einstein condensation of atomic gases has been  very
actively studied in recent years \cite{Fermi99,Dalfovo99}. Because
of the dilute character of the samples
 and the ability to control the atom-atom interactions, for the first time a detailed
 comparison with the theories of quantum  gases  has been made.
 Fermi gases, on the other hand, have only been
investigated
  experimentally for two years \cite{DeMarco99,Schreck01,Truscott01}.
   They are predicted to possess intriguing properties
  and may offer an interesting link with the behavior of electrons in metals and
   semi-conductors, and the possibility of
   Cooper pairing \cite{Stoof96} such as in high temperature superconductors
   and neutron stars.
 Mixtures of bosonic and fermionic quantum systems, with the
 prominent example of $^4$He- $^3$He\, fluids, have also stimulated
 intense theoretical and experimental activity \cite{Ebner70}. This has led to new physical
 effects including phase
separation, influence of the superfluidity of the Bose system on
the Fermi degeneracy and to new  applications
 such as the dilution refrigerator \cite{Ebner70,Moelmer98,Timmermans98}.

In this paper, we present a new mixture of bosonic and fermionic
 systems, a stable Bose-condensed gas immersed in a Fermi
sea. Confined in the same magnetic trap, both atomic species are in thermal equilibrium
with a temperature of $0.2(1)\,T_F\,<< T_{\rm C}$. The condensate is made of $^7$Li atoms
in internal state $ |F=1, m_F=-1\rangle$ while the degenerate Fermi gas is made of $^6$Li
atoms in $ |F=1/2, m_F=-1/2\rangle$ (fig.\,1). All previous experiments performed with
$^7$Li in $|F=2, m_F=2\rangle$ had condensate numbers limited to $N\leq 1400$ because of
the negative scattering length, $a=-1.4\,$nm, in this state \cite{Bradley97,Gerton00}. Our
condensate is produced in a state which has a positive, but small, scattering length,
$a=+0.27\,$nm \cite{Hulet}. The number of condensate atoms is typically $ 10^4$, and BEC
appears unambiguously both in the position distribution in the trap and in the standard
time of flight images. An interesting feature is the one-dimensional character of this
condensate, behaving as an ideal gas in the transverse direction of the
trap\cite{Petrov00,Goerlitz01}.

Because of the symmetrization postulate, colliding fermions have no {\it s}-wave
scattering at low energy and the {\it p}-wave contribution vanishes in the low temperature
domain of interest. Our method for producing simultaneous quantum degeneracy for both
isotopes
\begin{figure}[t]
\begin{center}
\epsfxsize=8cm \leavevmode \epsfbox{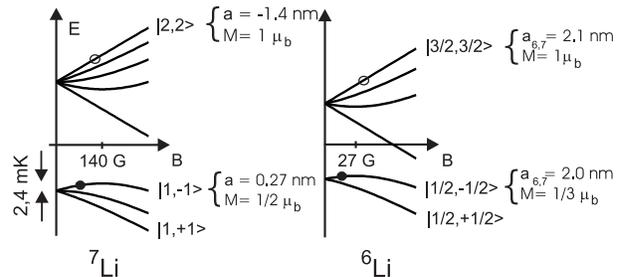} \caption{Energy
levels of $^7$Li and $^6$Li ground states in a magnetic field. Relevant scattering
lengths,a, and magnetic moments, M, are given. $\mu _b$ is the Bohr magneton. The
$|1,-1\rangle$ state (resp. $|1/2,-1/2\rangle$) is only trapped in fields weaker than 140
Gauss (resp. 27 Gauss). Open circles: first cooling stage; black circles: second cooling
stage.
\label{fig:LevelScheme}}
\end{center}
\end{figure}
of lithium is sympathetic cooling \cite{Schreck01,Truscott01}; {\it s}-wave collisions
between two different atomic isotopes are allowed and RF evaporation selectively removes
from the trap high energy atoms of one species. Elastic collisions subsequently restore
thermal equilibrium of the two-component gas at a lower temperature. Our experimental
setup has been described in detail in \cite{Schreck01,Mewes00}. A mixture of $^6$Li and
$^7$Li atoms is loaded from a magneto-optical trap into a strongly confining
Ioffe-Pritchard trap at a temperature of about $2\,$mK. As depicted in fig.1, this
relatively high temperature precludes direct magnetic trapping of the atoms in their lower
hyperfine state because of the shallow magnetic trap depth, 2.4\,mK for $^7$Li in $ |F=1,
m_F=-1\rangle$ and 0.2\,mK for $^6$Li in $|F=1/2, m_F=-1/2\rangle$. Therefore we proceed
in two steps. Both isotopes are first trapped and cooled in their upper hyperfine states.
There is no limitation for the trap depth and the confinement is better because the
magnetic moment is higher. Evaporation is performed selectively on $^7$Li using a
microwave field near $803$\, MHz that couples $ |F=2, m_F=2\rangle$ to $ |F=1,
m_F=1\rangle$. When both gases are cooled to a common temperature of about $9\,
\mu$K, atoms are transferred using a combination of microwave and
RF pulses into states $ |F=1, m_F=-1\rangle$ and $ |F=1/2, m_F=-1/2\rangle$ with an energy
far below their respective trap depths. Evaporative cooling is then resumed until the BEC
threshold is reached for $^7$Li.

\begin{figure}[t]
\begin{center}
\epsfxsize=8cm \leavevmode \epsfbox{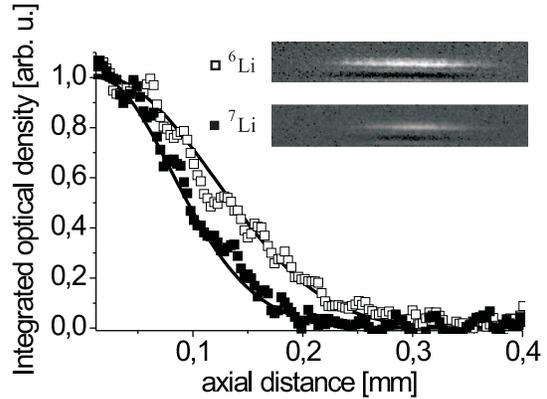} \caption{
\label{fig:FermiPressure} Observation of Fermi pressure.
Absorption images in the trap and spatial distributions integrated over the vertical
direction of $8.1\,10^4$ $^6$Li and $2.7\,10^4$ $^7$Li atoms in their higher hyperfine
states. The temperature is $1.4(1)\,\mu K$ corresponding to $T_{\rm C} $ for the bosons
and $0.33(5) T_{\rm F}$ for the fermions. Solid lines are the expected Bose and Fermi
distributions.}
\end{center}
\end{figure}

In the first series of experiments, both Li isotopes are
 trapped in their higher HF states. $^6$Li is
sympathetically cooled to Fermi degeneracy by performing 30 seconds of evaporative cooling
on $^7$Li \cite{Schreck01}. Trap frequencies for $^7$Li are $\omega_{\rm rad}=2
\pi*4000(10) s^{-1}$ and $\omega_{\rm ax}=2 \pi*75.0(1) s^{-1}$ with a bias field of
2\,G. Absorption images of both isotopes are recorded on a single CCD camera with a
resolution of $10\,\mu$m. Images are taken quasi-simultaneously (only 1\,ms apart) in the
trap or after a time of flight expansion. Probe beams have an intensity below saturation
and a common duration of $30\,\mu$s.
 Typical in-situ absorption images in the quantum  regime
 can be seen in fig.\ref{fig:FermiPressure}. Here the
temperature $T$ is $1.4(1)\,\mu$K and $T/T_{\rm F}=0.33(5)$, where the Fermi temperature
$T_{\rm F}$ is $(\hbar \bar{\omega}/k_{\rm B})(6 N_{\rm F})^{1/3}$, with $\bar{\omega}$
the geometric mean of the three oscillation frequencies in the trap and $N_{\rm F}$ the
number of fermions. For images recorded in the magnetic trap, the common temperature is
measured from the spatial extent of the bosonic cloud in the axial direction since the
shape of the Fermi cloud is much less sensitive to temperature changes when $T/T_{\rm
F}<1$\, \cite{Butts97}. The spatial distributions of bosons and fermions are recorded
after a 1 sec thermalization stage at the end of the evaporation. As the measured
thermalization time constant between the two gases, $0.15\,$s, is much shorter than
$1$\,s, the two clouds are in thermal equilibrium. Both isotopes experience the same
trapping potential. Thus the striking difference between the sizes of the Fermi and Bose
gases \cite{Truscott01} is a direct consequence of Fermi pressure. The measured axial
profiles in fig.\ref{fig:FermiPressure} are in excellent agreement with the calculated
ones (solid lines) for a Bose distribution at the critical temperature $T_{\rm C}$. In our
steepest traps, Fermi temperatures as high as $11\, \mu$K with a degeneracy of
$T/T_F=0.36$ are obtained. This $T_{\rm F}$ is a factor 3 larger than the single photon
recoil temperature at 671\,nm, opening interesting possibilities for light scattering
experiments \cite{Busch98}.

Our highest Fermi degeneracy in the $^6$Li $F=3/2$ state is
$T/T_{\rm F}=0.25(5)$ with $T_{\rm F}=4\,\mu$K, very similar to
 ref.\cite{Truscott01}. This limit is set by the fact that the boson temperature cannot be
lowered below $T_C$. Because of the negative scattering length in $^7$Li,
$|F=2,m=2\rangle$, the number of condensed atoms cannot exceed $\sim 300$ in our trap
\cite{Bradley97}. This, together with the condition that sympathetic cooling stops when
the heat capacity of the bosons approximately equals that of the fermions, limits the
Fermi degeneracy to about 0.3 \cite{Truscott01}.

In order to explore the behavior of a Fermi sea in the presence of a BEC with a
temperature well below $T_{\rm C}$, we perform another series of experiments with both
isotopes trapped in their lower HF state. As the $^7$Li scattering length is then positive
(fig.\ref{fig:LevelScheme}), a stable BEC with high atom numbers is now possible. To avoid
large dipolar relaxation, $^6$Li must also be in its lower HF state \cite{vanAbeelen97}.
First, sympathetic cooling is performed on the initial stretched states to $\sim 9\,\mu
K$.
 Then, to facilitate state transfer, the trap is adiabatically opened to frequencies
$\omega_{\rm rad}=2 \pi*100\, s^{-1}$ and $\omega_{\rm ax}=2 \pi*5\, s^{-1}$ (for $^7$Li,
F=2). The transfer of each isotope uses two microwave $\pi$ pulses. The first pulse at 803
MHz for $^7$Li (228 MHz for $^6$Li) transfers the atoms from $|2,2\rangle$ to
$|1,1\rangle$ ($|3/2,3/2\rangle$ to $|1/2,1/2\rangle$), a magnetically untrapped state
(see fig.\ref{fig:LevelScheme}). The second RF $\pi$ pulse at 1 MHz (1.3 MHz) transfers
them to $|1,-1\rangle$ ($|1/2,-1/2\rangle$) a magnetically trapped state. Adiabatic
opening of the trap cools the cloud, therefore decreasing the energy broadening of the
resonance and giving more time for the passage through untrapped states. The duration of
the $\pi$ pulses are $17\,\mu s$ and $13\,\mu s$ and more than $70\%$ of each isotope are
transferred. Finally the trap is adiabatically recompressed to the steepest confinement
giving $\omega_{\rm rad}=2 \pi*4970(10) s^{-1}$ and $\omega_{\rm ax}=2
\pi*83(1) s^{-1}$ for $^7$Li $|F=1, m=-1\rangle$, compensating for
the reduced magnetic moment.

\begin{figure}[t]
\begin{center}
\epsfxsize=8cm \leavevmode \epsfbox{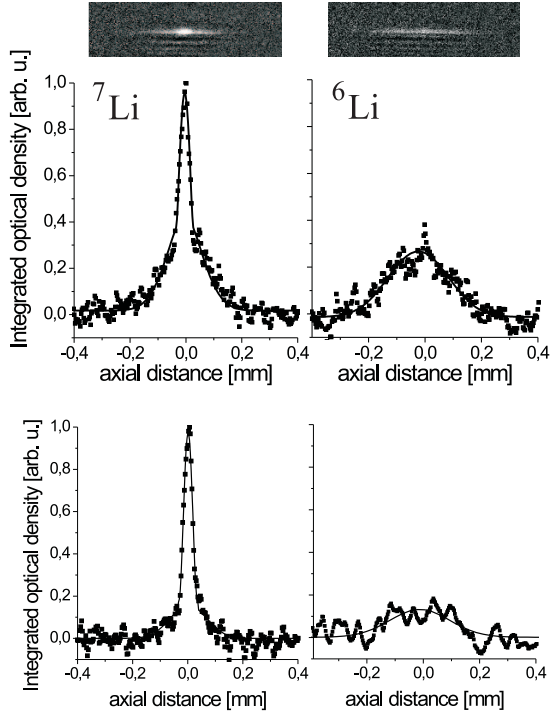}
\caption{\label{fig:BECAndFermiSea} Mixture
of Bose and Fermi gases. Top: In situ spatial distributions after sympathetic cooling with
$N_{\rm B}=3.5\,10^4$ and $N_{\rm F}=2.5\,10^4$.
 The Bose condensed peak ($8.5\,10^3$ atoms) is surrounded by
the thermal cloud which allows the determination of the common temperature. $T=1.6
\mu$K$=0.87\,T_{\rm C}=0.57\,T_{\rm F}$. The Fermi distribution is wider because of the
smaller magnetic moment and Fermi pressure. Bottom: profiles with a {\it quasi-pure}
condensate, with $N_{\rm B}=1.5\,10^4$, $N_{\rm F}=4\,10^3$. The barely detectable thermal
cloud indicates a temperature of $\simeq 0.28\,\mu$K $\simeq 0.2(1)\,T_{\rm
C}=0.2(1)\,T_{\rm F}$.}
\end{center}
\end{figure}

 Because of the very large reduction of the $^7$Li $s$-wave scattering cross section
(by a factor of 28) from the F=2 to the F=1 state, we were unable to reach runaway
evaporation with $^7$Li atoms alone in $F=1$. In contrast, the $^6$Li/$^7$Li cross section
is 28 times higher than the $^7$Li/$^7$Li one. We therefore use $^6$Li atoms as a buffer
gas to accelerate the thermalization rate of both gases. Two different methods were used
to perform the evaporation. The first consists in using two RF ramps on the HF transitions
of $^6$Li (from $|1/2,-1/2\rangle$ to $|3/2,-3/2\rangle$) and $^7$Li (from $|1,-1\rangle$
to $|2,-2\rangle$), which we balanced to maintain roughly equal numbers of both isotopes.
After 10\,s of evaporative cooling, Bose-Einstein condensation of $^7$Li occurs together
with a $^6$Li degenerate Fermi gas (fig.\ref{fig:BECAndFermiSea}). Surprisingly, a single
25 s ramp performed only on $^6$Li achieved the same results. In this case the equal
number condition was fulfilled because of the reduced lifetime of the $^7$Li cloud that we
attribute to dipolar collisional loss \cite{vanAbeelen97}.
 The duration of the RF
evaporation ramp was matched to this loss rate. In the following we concentrate on this
second, and simpler, evaporation scheme, sympathetic cooling of $^7$Li by evaporative
cooling of $^6$Li.

In fig.\ref{fig:BECAndFermiSea} in-situ absorption images of bosons and fermions at the
end of the evaporation are shown. The bosonic distribution shows the typical double
structure: a strong and narrow peak forms the condensate at the center, surrounded by a
much broader distribution, the thermal cloud. As the Fermi distribution is very
insensitive to temperature, this thermal cloud is a very useful tool for the determination
of the common temperature. Note that, as cooling was only performed on $^6$Li atoms, the
temperature measured on $^7$Li cannot be lower than the temperature of the fermions.
Measuring $N_{\rm B}$, $N_{\rm F}$, the condensate fraction $N_0/N_{\rm B}$, and
$\bar\omega$, we determine the quantum degeneracy of the Bose and Fermi gases. In
fig.\ref{fig:BECAndFermiSea}(top), the temperature is just below $T_{\rm C}$, $T=1.6\,
\mu$K$=0.87\,T_{\rm C}=0.57\,T_{\rm F}$. In fig.\ref{fig:BECAndFermiSea}(bottom) on the
contrary, the condensate is quasi-pure; $N_0/N_{\rm B}=0.77$; the thermal fraction is near
our detectivity limit, indicating a temperature of $\simeq 0.28\,\mu$K $\leq 0.2\,T_{\rm
C}=0.2\,T_{\rm F}$ with $N=8.2\, 10^3$
\begin{figure}[t]
\begin{center}
\epsfxsize=8cm \leavevmode \epsfbox{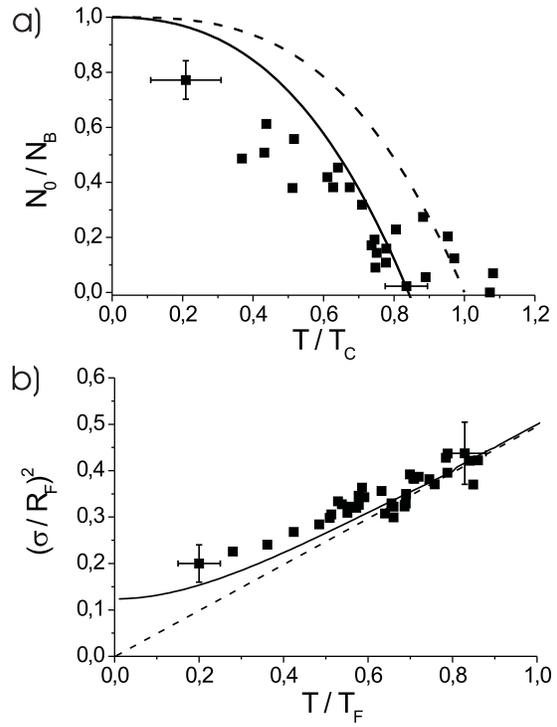}
\caption{Temperature dependence of mixtures of quantum gases: a) normalized BEC fraction as a
function of $T/T_{\rm C}$. Dashed line: theory in the thermodynamic limit. Solid line:
theory including finite size and trap anisotropy [2]; b) fermion cloud size: variance of
gaussian fit divided by the square of Fermi radius $R^{2}_{\rm F}= 2k_{\rm B}T_{\rm F}
/M\omega^{2}_{\rm ax}$ as a function of $T/T_{\rm F}$. Solid line:
theory. Dashed line: Boltzmann gas.
\label{fig:BECfraction}}
\end{center}
\end{figure}
bosons and $4\, 10^3$ fermions.
 Clearly a more
sensitive thermal probe is required now to investigate the temperature domain $T<
0.2\,T_{\rm F}$. An elegant method uses the measurement of thermalization rates with
impurity atoms including Pauli blocking \cite{Ferrari99,DeMarco01}. The condensate
fraction $N_0/N_{\rm B}$ as a function of $T/T_{\rm C}$ is shown in
fig.\ref{fig:BECfraction} (a), while the size of the fermi gas as a function of $T/T_{\rm
F}$ is shown in fig.\ref{fig:BECfraction} (b). With the strong anisotropy ($\omega_{\rm
rad}/\omega_{\rm ax}=59$) of our trap, the theory including anisotropy and finite number
effects differs significantly from the thermodynamic limit \cite{Dalfovo99}, in agreement
with our measurements even though there is a $20\%$ systematic uncertainty on our
determination of $T_{\rm C}$ and $T_{\rm F}$.

\begin{figure}[t]
\begin{center}
\epsfxsize=8cm \leavevmode \epsfbox{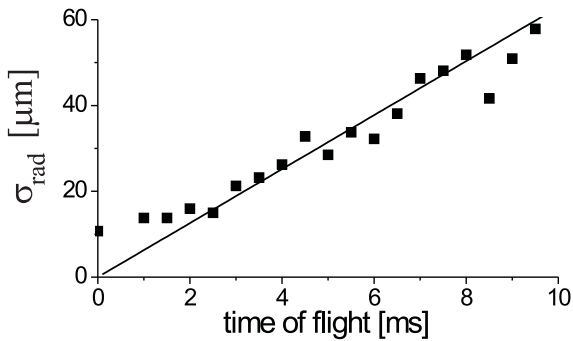} \caption{Signature
of 1D condensate. Radial size of expanding condensates with $10^4$
atoms as a function of time of flight. The straight line is the
expected behavior for the  expansion of the ground state radial
harmonic oscillator. \label{fig:tof} }
\end{center}
\end{figure}

Because of the small scattering length, this $^7$Li condensate has
 interesting properties. The
 time of flight images, performed after expansion times of 0-10\,ms
with $N_{\rm 0}=10^4$ condensed atoms,
 reveal that the condensate is a one-dimensional (1D) condensate.
In contrast to typical condensates in the Thomas-Fermi (TF) regime, where the release of
interaction energy leads to a fast increase in radial size, our measurements agree to
better than 5\% with the time development of the radial ground state wave function in the
harmonic magnetic trap (fig.\ref{fig:tof}). This behavior is
expected when the chemical
potential $\mu$ satisfies the condition $\mu < \hbar \omega_{\rm rad}$. Searching for the
ground state energy of the many-body system with a Gaussian ansatz radially and TF shape
axially \cite{Petrov00}, we find that the mean-field interaction increases the size of the
Gaussian by $\approx 3\%$. The calculated TF radius is $28\,\mu$m or 7 times the axial
harmonic oscillator size and is in good agreement with the measured radius, $30\,\mu$m in
fig. \ref{fig:BECAndFermiSea}. Thus with $\mu =0.45\, \hbar \omega_{\rm rad}$, the gas is
described as an ideal gas radially but is in the TF regime axially. This 1D situation was
also realized recently in sodium condensates \cite{Goerlitz01}. As $\mu <
\hbar
\omega_{\rm rad}$ implies that the linear density of a 1D
condensate is limited to $\simeq 1/a$, the 1D regime is much
easier to reach with $^7$Li (small $a$) than with Na or $^{87}$Rb
which have much larger scattering lengths.

What are the limits of this BEC-Fermi gas cooling scheme? First, the $1/e$ condensate
lifetime of about 3\,s in this steep trap will limit the available BEC-Fermi gas
interaction time. Second, the boson-fermion mean field interaction can induce a spatial
phase separation \cite{Moelmer98} that prevents thermal contact between $^7$Li and $^6$Li.
Using the method of \cite{Moelmer98} developed for $T=0$, we expect, for the parameters of
fig.\ref{fig:BECAndFermiSea} (top), that the density of fermions is only very slightly
modified by the presence of the condensate in accordance with our observations. Third,
because of the superfluidity of the condensate, impurity atoms (such as $^6$Li), which
move through the BEC slower than the sound velocity $v_{\rm c}$, are no longer scattered
\cite{Timmermans98,Chikkatur00}. When the Fermi velocity $v_{\rm F}$ becomes smaller than
$ v_{\rm c} $, cooling occurs only through collisions with the bosonic thermal cloud, thus
slowing down drastically. With $10^4$ condensed atoms, the velocity is $\simeq 0.9\,$cm/s,
corresponding to a limiting temperature of about 100\,nK, a factor 3 lower than our
currently measured temperature.

In summary, we have produced a new mixture of  Bose and Fermi
quantum gases. Future work will explore the degeneracy limits of
this mixture with the sensitive temperature probe mentioned above
\cite{Ferrari99}. Phase fluctuations of the 1D $^{7}$Li condensate
should also be detectable via density fluctuations in time of
flight images, as recently reported in \cite{Dettmer01}. The
transfer of the BEC into $|F=2, m=2\rangle$ with negative $a$
should allow the production of bright solitons and of large
unstable condensates  where interesting and still unexplained
dynamics  has been recently observed \cite{Gerton00,Roberts01}.
Finally, the large effective attractive interaction between $^6$Li
$|F=1/2, m_F=+1/2\rangle$ and $|F=1/2, m_F=-1/2\rangle$ makes this
atom an attractive candidate for searching for  BCS pairing if the
temperature can be made sufficiently low \cite{Stoof96}.

 We are grateful  to Y. Castin, J. Dalibard, C.
Cohen-Tannoudji, and G. Shlyapnikov for useful discussions.
F.\,S., and K.\,C. were supported by
 a fellowship from the DAAD and by MENRT. This work was
  supported by CNRS and Coll\`ege de France. Laboratoire Kastler Brossel is
{\it Unit\'e de recherche de l'Ecole Normale Sup\'erieure et de
l'Universit\'e Pierre et Marie Curie, associ\'ee au CNRS}.

\end{document}